\documentclass[prl,twoside,showpacs,superscriptaddress,twocolumn]{revtex4}

\usepackage{latexsym}
\usepackage{amssymb}

\usepackage{graphicx}

\begin{document}

\title{Fibre-optics implementation of asymmetric phase-covariant quantum cloner}

\author{Lucie Bart{\accent23 u}\v{s}kov\'{a}}
\affiliation{Department of Optics, Palack\'y University,
     17.~listopadu 50, 772\,00 Olomouc, Czech~Republic}

\author{Miloslav Du\v{s}ek}
\affiliation{Department of Optics, Palack\'y University,
     17.~listopadu 50, 772\,00 Olomouc, Czech~Republic}

\author{Anton\'{\i}n \v{C}ernoch}
\affiliation{Department of Optics, Palack\'y University,
     17.~listopadu 50, 772\,00 Olomouc, Czech~Republic}
\affiliation{Joint Laboratory of Optics of Palack\'{y} University and
     Institute of Physics of Academy of Sciences of the Czech Republic,
     17. listopadu 50A, 772\,00 Olomouc, Czech Republic}

\author{Jan Soubusta}
\affiliation{Joint Laboratory of Optics of Palack\'{y} University and
     Institute of Physics of Academy of Sciences of the Czech Republic,
     17. listopadu 50A, 772\,00 Olomouc, Czech Republic}

\author{Jarom{\'\i}r Fiur\'{a}\v{s}ek}
\affiliation{Department of Optics, Palack\'y University,
     17.~listopadu 50, 772\,00 Olomouc, Czech~Republic}

\date{\today}

\begin{abstract}

We present the experimental realization of optimal
symmetric and asymmetric  phase-covariant
$1\to2$ cloning of qubit states using fiber optics. State
of each qubit is encoded into a single photon which can
propagate through two optical fibers. The operation
of our device is based on one- and two-photon interference.
We have demonstrated creation of two copies of any state of
a qubit from the equator of the Bloch sphere. The
measured fidelities of both copies are close to the
theoretical values and they surpass the theoretical
maximum obtainable with the universal cloner.

\end{abstract}


\maketitle



The quantum no-cloning theorem \cite{Wootters82} lies at the heart of quantum
information theory. The apparently simple observation that perfect copying
of unknown quantum states is impossible has  profound
consequences. On the fundamental side, it prevents superluminal communication with entangled states,
thereby guaranteeing the peaceful coexistence of quantum mechanics and theory of relativity.
On the practical side, this theorem is behind the security of the quantum key
distribution schemes which rely on the fact that any attempt to measure or
copy an unknown quantum state results in the disturbance of this state.
Going beyond the no-cloning theorem,
Bu\v{z}ek and Hillery in a seminal paper introduced the concept of the universal
approximate quantum cloning machine that optimally approximates the
forbidden transformation $|\psi\rangle \rightarrow |\psi\rangle |\psi\rangle$
\cite{Buzek96}.
Today, optimal quantum cloners are known for many different cases and scenarios
\cite{Scarani05,Cerf06}.
During recent years, growing attention has been paid to the experimental implementation of
quantum cloning machines and, in particular,  optimal cloning of polarization states of single
photons via stimulated parametric downconversion or via photon bunching on a beam
splitter has been successfully demonstrated
\cite{Lamas-Linares02,DeMartini04,Ricci04,Irvine04,Khan04,Sciarrino05}.

Besides giving an insight into the fundamental limits on distribution of quantum
information, the quantum cloning machines turned out to be
very efficient eavesdropping attacks on the quantum key distribution protocols
\cite{Fuchs97,Cerf02,Bruss02,Dusek06}. In
this context one is particularly interested in the \emph{asymmetric} quantum
cloners that produce two copies with different fidelities.  In this way, the
eavesdropper can control the trade-off between the information gained on a secret
cryptographic key and the amount of noise added to the copy which is sent down the
channel to the authorized receiver. While the theory of optimal asymmetric quantum copying
machines is well established (see, e.g. the recent reviews \cite{Scarani05,Cerf06}), the experimental optical realization of such machines
has received considerably less attention. This might be attributed to the fact that the
asymmetric cloning operations exhibit much less symmetry than the corresponding
symmetric ones. To the best of our knowledge, asymmetric quantum cloning
of single-photon states has been so far achieved only in a single experiment, where
universal asymmetric copying of polarization states was performed by means of partial
quantum teleportation \cite{Zhao05}.

In this Letter, we report on the experimental implementation of the optimal
$1 \rightarrow 2$ phase-covariant asymmetric cloning of photonic qubits represented by a single
photon propagating in two single-mode optical fibers. The phase-covariant copying
machine optimally clones all states on the equator of the Bloch sphere,
$|\psi\rangle=\frac{1}{\sqrt{2}}(|0\rangle+e^{i\phi}|1\rangle)$.
Our experiment is
based on the interplay of single- and two-photon interference of two photons in an
 optical network built from optical fibers. This approach
has several important technological advantages. First, the single-mode fibers
guarantee very high interference visibility. Second,  variable ratio couplers
enable to easily change in a controlled way the cloning transformation and we are
thus able to demonstrate the whole class of the optimal asymmetric cloners. In contrast
to our previous experiment on the optimal symmetric phase-covariant cloning of
polarization states of single photons \cite{Cernoch06}, with the present fiber-based 
scheme \cite{Bartuskova06} we are able to achieve fidelities exceeding the limit of optimal universal cloning
machine. This is rather challenging because the fidelities of the
optimal  universal and phase-covariant cloners are very close. For instance, for a
symmetric cloner we have $F_{\mathrm{univ}}=\frac{5}{6} \approx 0.833$ and
$F_{\mathrm{pc}}=\frac{1}{2}(1+\frac{1}{\sqrt{2}})\approx 0.854$ so the fidelities
differ only by  $2.1$\%.

\begin{figure}
\centerline{\includegraphics[width=\linewidth]{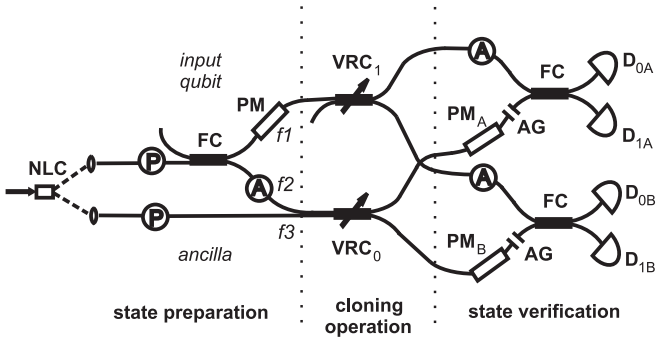}}
\caption{Experimental setup. NLC denotes nonlinear crystal,
P polarizers, FC fiber couplers, A attenuators, PM phase
modulators, VRC variable-ratio couplers, AG adjustable
air-gaps, D detectors.} \label{setup}
\end{figure}

 The optimal asymmetric phase-covariant cloning
transformation requires only a single blank copy in addition to the input qubit to be cloned and reads
\cite{Niu99},
\begin{eqnarray}
|0\rangle&\rightarrow & |00\rangle,  \nonumber \\
 |1\rangle &\rightarrow & \sqrt{q} \, |10\rangle + \sqrt{1-q} \, |01\rangle,
\label{pc_cloning}
\end{eqnarray}
where $q \in [0, 1]$ characterizes the asymmetry of the clones and for the symmetric
cloner $q=\frac{1}{2}$. The fidelities of the two clones are given by
\begin{equation}
  F_A= \frac{1}{2} \left( 1+ \sqrt{q} \right), \quad
  F_B= \frac{1}{2} \left( 1+ \sqrt{1-q} \right).
 \label{fid-asym}
\end{equation}
In our scheme (see Fig.~\ref{setup}) each qubit is represented by
a single photon which may propagate in two optical fibers
and the basis states $|0\rangle$ and $|1\rangle$ correspond to the presence 
of the  photon in the first or second fiber, respectively. The state of ancilla 
photon is initially $|0\rangle$ while the signal photon can be prepared in
an arbitrary state from the equator of the Bloch sphere.
The two photons impinge on two unbalanced beam splitters
(variable ratio couplers VRC$_0$ and
VRC$_1$) with different splitting ratios. Let us suppose that real
amplitude transmittances and reflectances of VRC$_0$ and VRC$_1$ are
$t_0$, $r_0$ and $t_1$, $r_1$, respectively. We use the notation
$R_j=r_j^2$ and $T_j=t_j^2$ for the intensity reflectances and
transmittances and $R_j+T_j=1$ for a lossless beam splitter. 
In the experiment, we accept only the events when there
is a single photon detected in each output pair of fibers corresponding to the
clone A and B, respectively. The cloning transformation is thus implemented
conditionally, similarly to other optical cloning experiments. The
resulting conditional transformation reads \cite{Fiurasek03}
\begin{eqnarray}
|0\rangle_{\mathrm{Sig}}|0\rangle_{\mathrm{Anc}} & \rightarrow & (r_0^2-t_0^2) \, |00\rangle, \nonumber \\[2mm]
|1\rangle_{\mathrm{Sig}}|0\rangle_{\mathrm{Anc}} & \rightarrow & 
r_0 r_1 \, |10\rangle - t_0 t_1 \, |01\rangle.
\label{HV_transf}
\end{eqnarray}
This  becomes equivalent to the optimal cloning operation (\ref{pc_cloning}) up to a constant
prefactor representing the probability amplitude of successful cloning,
if the following equations hold,
\[
r_0 r_1=\sqrt{q}(r_0^2-t_0^2), \qquad t_0t_1 = -\sqrt{1-q}(r_0^2-t_0^2).
\]
Taking the square of the ratio of these two equations, we arrive at
\begin{equation}
  R_1 = \frac{q(1-R_0)}{q(1-R_0)+(1-q)R_0},
 \label{R2-asym}
\end{equation}
and from the normalization $T_1+R_1=1$ we find after some algebra
that $R_0$ can be determined as a root of a cubic polynomial,
\begin{equation}
  R_0 (1-R_0) + \left[ R_0 (2q-1) - q \right] (2 R_0 -1)^2 = 0.
 \label{R1-asym}
\end{equation}
The resulting reflectances are given in  Tab.~\ref{tab-asym} for several values of the
asymmetry parameter $q$. The equations
have always two physically significant solutions that also
require different signs of amplitude reflectances and
transmittances. We have always selected the ``less
unbalanced'' splitting ratios as they are more convenient
from the experimental point of view.

Our experimental setup is shown in Fig.~\ref{setup}.
A pair of signal and ancilla photons is prepared by means of
frequency-degenerate type-I
spontaneous parametric down-conversion in a 10-mm-long
$\mbox{LiIO}_3$ nonlinear crystal pumped by a krypton-ion
cw laser (413.1 nm), similarly as in our previous experiments
\cite{Cernoch06,Bartuskova06}. The signal photon is split by a
fiber coupler FC into two fibers. The basis states of the
signal qubit, $|0\rangle$ and $|1\rangle$, correspond to the
presence of a photon either in fiber $f2$ or $f1$,
respectively. The intensity ratio and phase difference
between these two modes determine the input state of the
signal qubit. Preparation of the state is affected by unequal losses in the 
two optical paths $f1$ and $f2$ which alter the effective splitting ratio of FC.
This effective splitting ratio is measured with the help of a semiconductor 
laser and a PIN photodiode and the attenuator in mode $f2$ is adjusted 
in such a way that the setup is balanced and at the end of the state preparation block 
the signal photon is evenly split between  $f1$ and $f2$. 
Various equatorial
qubit states $\frac{1}{\sqrt{2}}(|0\rangle+e^{i\phi}|1\rangle)$  can be then prepared 
by changing only the voltage applied to the phase modulator PM which sets the 
relative phase $\phi$.  The ancilla is in a fixed state $|0\rangle$ which 
corresponds to a single photon propagating through the fiber $f3$.

The cloning operation is realized by
two variable-ratio couplers VRC$_0$ and VRC$_1$. VRC$_0$ forms the core of
Hong-Ou-Mandel (HOM) interferometer \cite{HOM}. For optimal cloning
it is necessary to achieve precise time overlap of the two photons at VRC$_0$ and match
their polarizations. To accomplish these tasks the
splitting ratio of VRC$_0$ is set to 50:50. We typically
reach visibilities of HOM dip around $98\%$. 
Then  the VRC$_0$ splitting ratio is changed to the required value depending on the
asymmetry parameter $q$, c.f. Table I.

The two fiber-based Mach-Zehnder (MZ) interferometers are
adjusted using only the signal beam from the nonlinear
crystal, the ancilla beam is blocked. Detection rates at 
each detector are measured and used for the alignment of the setup.
 First the intensity transmittances of the whole arms of
each MZ interferometer are balanced with the help of the attenuators 
in the ``state verification'' part of the setup, which compensates for 
the unequal losses caused by the splitting ratios of variable ratio
couplers, the phase modulators, air-gaps and other factors. 
Visibilities are maximized by
adjusting zero path differences and aligning polarizations
in the interferometers. In this setting visibilities 
above $97\%$ are achieved.  After this step we unbalance the MZ interferometers
properly again:
From the transmittances and reflectances of VRC$_0$ and VRC$_1$ 
used in the experiment and given in Table I  we can determine what should be the 
detection rates for equal  losses in the  optical paths from VRC$_0$ and VRC$_1$ 
to FC.  So, we tune the attenuators until we reach the point 
where these  optical-path losses  are balanced. This ensures that each detection 
block performs projections onto  the states on the equator of the Bloch sphere.


\begin{table}
\begin{tabular}{|c||c|c||c|c||c|c|}
\hline
\multicolumn{5}{|c||}{Theory} & \multicolumn{2}{c|}{Experiment} \\
\hline
 $q$ & $R_0$ &  $R_1$ &  $F_A$ &  $F_B$ &  $F_A$ & $F_B$ \\
\hline
0.5  &  0.789  &  0.211  &  0.854  &  0.854  &  $0.854 \pm 0.004$ &  $0.834 \pm 0.004$  \\
0.6  &  0.801  &  0.271  &  0.887  &  0.816  &  $0.881 \pm 0.006$ &  $0.789 \pm 0.005$   \\
0.7  &  0.817  &  0.344  &  0.918  &  0.774  &  $0.905 \pm 0.003$ &  $0.754 \pm 0.005$  \\
0.8  &  0.838  &  0.436  &  0.947  &  0.724  &  $0.935 \pm 0.002$ &  $0.714 \pm 0.006$  \\
0.9  &  0.872  &  0.570  &  0.974  &  0.658  &  $0.964 \pm 0.002$ &  $0.641 \pm 0.004$  \\
1.0  &  1.000  &  1.000  &  1.000  &  0.500  &   ---   &   ---   \\
\hline
\end{tabular}
\caption{Asymmetric phase covariant cloner. Table shows calculated
         reflectances of variable ratio couplers and theoretical
         and measured fidelities for different parameters of asymmetry $q$.
         For $q<0.5$ the clones are just interchanged.
         Error intervals represent statistical errors.}
\label{tab-asym}
\end{table}

To reduce the effect of a phase drift between arms of each
MZ interferometer caused by fluctuations of temperature and
temperature gradients we apply both passive and active
stabilization. The experimental setup is thermally isolated
in a polystyrene box. After this precaution the phase drift
in each MZ interferometer has the average value $\pi/1000$
per second. Therefore the active stabilization of phase
differences is repeatedly applied after each three-second
measurement period \cite{Bartuskova06}. Only one beam from the crystal is used
for the active stabilization and the other one is blocked.
In each stabilization cycle the values of the phase drifts are
estimated and they are compensated by  means of phase
modulators PM$_A$ and PM$_B$. In this way, both interferometers 
are stabilized simultaneously. 

We have experimentally realized cloning operation for the
five values of asymmetry parameter $q$ shown in
Tab.~\ref{tab-asym}. For each $q$ related to
given splitting ratios of the couplers VRC$_0$ and VRC$_1$,
various states from the equator of the Bloch sphere
were cloned. Two detection blocks are used to  measure simultaneously
fidelities of both clones. 
Each block consists of an attenuator, a
phase modulator, a 50:50 fiber coupler and two detectors
(Perkin-Elmer single-photon counting modules employing
silicon avalanche photodiodes with quantum efficiency
$\eta\approx 50\%$). The cloning is successful only if one
photon passes to the modes of qubit A and the other one to
the modes of qubit B. Hence coincidences between detectors
$D_{iA}$ and $D_{jB}$ ($i,j=0,1$) are counted. The signals
from detectors are processed by coincidence electronics
based on time-to-amplitude convertors and single-channel
analyzers with a two-nanosecond coincidence window.

\begin{figure}
\centerline{
\includegraphics[width=0.95\linewidth]{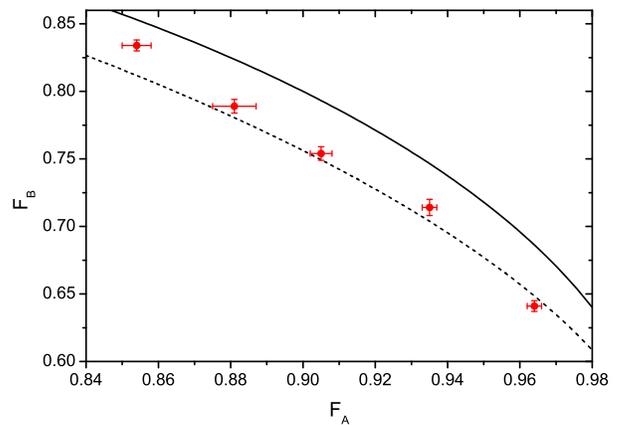}}
\caption{Dependence of fidelity $F_B$ on $F_A$ for
phase-covariant cloner with different asymmetries. Symbols
denote experimental data, solid line represents theoretical
prediction for optimal asymmetric phase-covariant cloner.
Dashed line shows theoretical prediction for optimal
asymmetric universal cloner.} \label{asym}
\end{figure}


The measurement basis for each clone consists of the input signal
state and the state orthogonal to it, which is guaranteed by the 
alignment procedure described above.  Four
coincidence rates $C^{++}$, $C^{--}$, $C^{+-}$ and $C^{-+}$
were measured. The first sign concerns clone A and the other
one clone B; ``$+$'' means projection to the original signal
state and ``$-$'' to its orthogonal complement. Fidelities
of clones read
\begin{eqnarray}
F_A=\frac{C^{++}+C^{+-}}{C^{++}+C^{--}+C^{+-}+C^{-+}}, \nonumber \\[2mm]
F_B=\frac{C^{++}+C^{-+}}{C^{++}+C^{--}+C^{+-}+C^{-+}}.
\end{eqnarray}

Our results are summarized in Fig.~\ref{asym} and in
Tab.~\ref{tab-asym}. The fidelities for each value of
asymmetry are averaged over all cloned signal states from
the equator. Fig.~\ref{asym} shows the fidelity of the
second clone as a function of the fidelity of the first
clone. One can see that there is a small systematic error --
measured fidelities are always $1-2\%$ lower than their
theoretical values. This is caused by misalignments,
limited precision of parameter setting and a phase drift in
both MZ interferometers during the measurement period.
However, the qualitative agreement between the theoretical
curve for the optimal asymmetric phase-covariant cloner, determined 
by Eq. (\ref{fid-asym}), and the measured data is very good. For comparison, 
the dashed line indicates the trade-off between the fidelities of the optimal 
universal asymmetric cloner \cite{Cerf06},
\[
F_A=1-\frac{(1-p)^2}{2(1-p+p^2)}, \qquad
F_B=1-\frac{p^2}{2(1-p+p^2)},
\]
where $p\in[0,1]$. Note that most of the experimental points lie in
the area inaccessible by any universal cloning machine.

\begin{figure}
\centerline{
\includegraphics[width=0.95\linewidth]{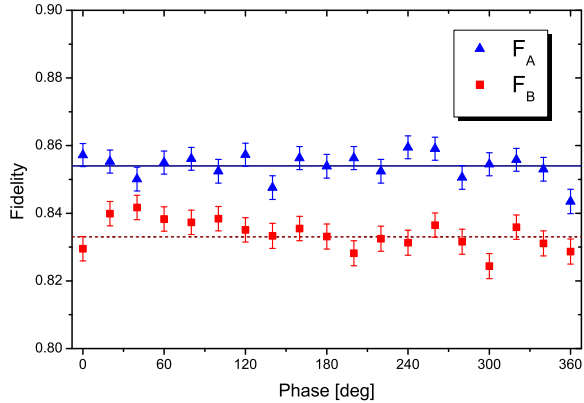}}
\caption{Symmetric phase-covariant cloner. Fidelities
$F_A$  and $F_B$  are plotted 
as functions of input-state phase $\phi$.
Symbols denote experimental data, solid line represents
theoretical prediction for the phase-covariant cloner, and
the dashed line shows theoretical prediction for the
universal cloner. Error bars represent statistical errors.}
\label{sym}
\end{figure}

As an example, Fig.~\ref{sym} shows data measured for the symmetric
phase-covariant cloner ($q=0.5$). The splitting ratio of
VRC$_0$ was set to 21:79 whereas the splitting ratio of VRC$_1$
to 79:21. The measurement was done for phases from
$0^\circ$ to $360^\circ$ with a step of $20^\circ$. For each
phase 40 three-second measurements were performed.
Displayed fidelities are calculated from data measured simultaneously by
all four detectors. The  unequal detector efficiencies were compensated 
by proper rescaling of the measured coincidences. As
expected, fidelities are nearly independent on phase. 
We can see that fidelities  $F_A$ and $F_B$ of symmetric cloner 
are in fact slightly different due to imperfections of our setup.
The splitting ratio of VRC$_0$ was
always set in such a way that the greater part of ancilla went
to clone B. Therefore the visibility of HOM dip lower than
$100\%$ and the inaccuracy of position setting in HOM dip
have stronger influence on fidelity $F_B$ than $F_A$.
However, the average of fidelities $F_A$ and $F_B$
overcomes the bound for universal cloner.


Because none of the output fiber couplers of MZ
interferometers is precisely 50:50 the visibility of single
photon interference cannot be perfect at both output ports
of the coupler \cite{HDH96}. Therefore we have also
measured all four coincidence rates sequentially at only
one pair of detectors using proper phase shifts at phase
modulators PM$_A$ and PM$_B$. We had chosen the two
detectors where the visibilities were maximized. Using only
these two detectors we have obtained fidelities (averaged
over all phases): $F_A=0.840\pm0.009$, $F_B=0.850\pm0.009$.
In this kind of measurement no compensation for
different detector efficiencies was needed.

In summary, we have demonstrated optimal symmetric and asymmetric phase-covariant 
cloning of single-photon states. Using  fiber optics  allowed us to reach
very high visibilities and achieve fidelities exceeding the maximum obtainable by any
universal cloning machine. Our implementation is compatible with fiber-based
communication systems and represents a promising platform for realization of various
protocols for quantum information processing.


This research was
supported by the projects LC06007, 1M06002 and MSM6198959213 of the
Ministry of Education of the Czech Republic and by the
SECOQC project of the EC (IST-2002-506813).


\end{document}